# Strong and tunable spin-orbit interaction in a single crystalline InSb nanosheet

Yuanjie Chen[1], Shaoyun Huang[1], Dong Pan[2,3], Jianhong Xue[1], Li Zhang[1], Jianhua Zhao[2,3,†], and H. Q. Xu[1,3,*]

[1]*Beijing Key Laboratory of Quantum Devices, Key Laboratory for the Physics and Chemistry of Nanodevices and Department of Electronics, Peking University, Beijing 100871, China*

[2]*State Key Laboratory of Superlattices and Microstructures, Institute of Semiconductors, Chinese Academy of Sciences, P.O. Box 912, Beijing 100083, China*

[3]*Beijing Academy of Quantum Information Sciences, Beijing 100193, China*

Corresponding authors. Emails: *hqxu@pku.edu.cn; †jhzhao@semi.ac.cn

(Dated: November 16, 2020)

**ABSTRACT**

A dual-gate InSb nanosheet field-effect device is realized and is used to investigate the physical origin and the controllability of the spin-orbit interaction in a narrow bandgap semiconductor InSb nanosheet. We demonstrate that by applying a voltage over the dual gate, efficiently tuning of the spin-orbit interaction in the InSb nanosheet can be achieved. We also find the presence of an intrinsic spin-orbit interaction in the InSb nanosheet at zero dual-gate voltage and identify its physical origin as a build-in asymmetry in the device layer structure. Having a strong and controllable spin-orbit interaction in an InSb nanosheet could simplify the design and realization of spintronic deceives, spin-based quantum devices and topological quantum devices.

# Introduction

    Low-dimensional narrow bandgap InSb nanostructures, such as nanowires and quantum wells, have in recent years attracted great interests. Due to their small electron effective mass, strong spin-orbit interaction (SOI), and large Landé g-factor, these nanostructures have potential applications in high-speed electronics[1], infrared optoelectronics[2], spintronics[3], quantum electronics[4,5] and topological quantum computation[6]. The past decade has witnessed booming investigations of devices made from epitaxially grown InSb nanowires, including field-effect transistors[7,8], single[9-11] and double quantum dots[12,13], and semiconductor-superconductor hybrid quantum devices[14-17]. Among the most influential, pioneer developments are the topological superconducting quantum devices made from InSb nanowires[14,16], in which zero-energy modes, a signature of Majorana fermions[18,19] in solid state, were detected and studied. However, to build a device in which braiding of topological quantum states, such as Majorana fermions, can be conveniently performed and thus topological quantum computations can be designed and realized, it could be inevitable to move from single-nanowire structures to multiple-nanowire[20,21] and two-dimensional (2D) planar quantum structures[22-24]. Recently, high-quality InSb/InAlSb heterostructured quantum wells[25,26] and free-standing InSb nanosheets[27-30] have been achieved by epitaxial growth techniques. In comparison with InSb/InAlSb quantum well systems, the free-standing InSb nanosheets have advantages in direct contact by metals, including superconducting materials, in easy transfer to different substrates, and in convenient fabrication of dual-gate structures. With use of free-standing InSb nanosheets, lateral quantum devices, such as planar quantum dots[31] and superconducting Josephson junctions[32-34] have been successfully fabricated. A most intriguing perspective of these layered materials is to build topological superconducting structures from them, in which Majorana fermions and parafermions[35,36] can be created and manipulated, enabling a different route of developments towards topological quantum computation technology. A desired ingredient in constructing topological superconducting states from such a semiconductor nanostructure is strong SOI (with a few 100 nm or shorter in spin-orbit length and about 100 µeV or larger in spin-orbit energy) possessed in the material[37,38]. Comprehensive studies of SOI have been carried out for InSb nanowires[39,40] and quantum wells[41]. However, a desired study of SOI and, in particular, its controllability has not yet been carried out for free-standing InSb nanosheets, although it is highly anticipated that such a study would lead to great advancement in the developments of spintronics, quantum-dot based spin-orbit qubits, and topological quantum computation technology.

    In this article, we report on magnetotransport measurements of an epitaxially-grown,

free-standing, zincblende InSb nanosheet and on employment of dual-gate technique to achieve tunable SOI in the nanosheet. Key electron transport characteristic lengths, such as the mean free path, phase coherence length and SOI length, in the nanosheet are extracted from the measurements of the low-field magnetoconductance. We show that a strong SOI is present in the InSb nanosheet and is greatly tunable using a voltage applied over the dual gate. We also demonstrate, through band diagram simulation for the experimental structure setups, that the origin of an intrinsic SOI observed in the InSb nanosheet at zero dual-gate voltage comes from the build-in structure asymmetry in the dual-gate device. The advancement made in this work in understanding and controlling of strong SOI in the InSb nanosheet will greatly simply the design and implementation technology for the construction of spintronic devices, spin-orbit qubits, and topological quantum devices.

## Results and discussion

Dual-gate InSb nanosheet device

The dual-gate device studied in this work is made from a free-standing, single-crystalline, zincblende InSb nanosheet on an *n*-doped silicon (Si) substrate covered by a 300-nm-thick layer of silicon dioxide ($SiO_2$) on top, using standard nanofabrication techniques (see Methods). Figure 1a shows a scanning electron microscope (SEM) image of the device and the measurement circuit setup. Figure 1b shows a schematic view of the layer structure of the device. The InSb nanosheet in the device is contacted by four stripes of Ti/Au (contact electrodes). The *n*-doped Si substrate (contacted by a thin gold film at the bottom) and the $SiO_2$ layer are employed as the bottom gate and the gate dielectric. The top gate is made from a Ti/Au film with a layer of Hafnium dioxide ($HfO_2$) as the top gate dielectric. The nanosheet has a width of ~550 nm and a thickness of ~30 nm (estimated based on the calibrated contrast in the SEM image). The separation between the two inner Ti/Au electrodes is 1.1 μm. Low-temperature transport measurements of the dual-gate device is carried out in a physical property measurement system (PPMS) cryostat, equipped with a uniaxial magnet, in a four-probe configuration, in which a 17 Hz AC excitation current ($I$) of 100 nA is supplied through the two outer electrodes and the voltage drop ($V$) between the two inner contact electrodes is recorded. The nanosheet channel conductance $G$ is obtained from $G = I/V$. In measurements for the magetoconductance, $\Delta G = G(B) - G(B = 0)$, the magnetic field is applied perpendicular to the InSb nanosheet plane.

Figure 1c shows the measured conductance of the InSb nanosheet in the device as a function of voltages, $V_{BG}$ and $V_{TG}$, applied to the bottom and top gates (transfer characteristics). Figure 1d shows a horizontal line cut of Fig. 1c (bottom gate transfer characteristics) at $V_{TG} = 0$ V, while Fig. 1e shows a vertical line cut of Fig. 1c (top gate transfer characteristics) at $V_{BG} = 0$ V. Conductance fluctuations superimposed on the transfer curves are observable. These fluctuations are reproducible and arise from universal conductance fluctuations[42] (UCF). Overall, the top gate shows a strong coupling to the InSb nanosheet, while the bottom gate shows a relatively weak coupling to the nanosheet. The former is in accordance with the fact that a short distance between the top gate and the nanosheet and a high dielectric material (HfO$_2$ in this case) are employed in the device. From Fig. 1b, one can infer that an electric field stretching perpendicularly through the InSb nanosheet can be present and can be tuned by a voltage applied over the two gates (dual-gate voltage).

The carrier density in the InSb nanosheet can be estimated from the measured transfer characteristics. Here, we extract the carrier density, at a fixed top gate voltage of $V_{TG} = 0$ V, from $n = C_{gs} \times \frac{V_{BG} - V_{BG}^{th}}{e}$, where $e$ denotes the elementary charge and $C_{gs} = \frac{\varepsilon \varepsilon_0}{d}$ is the unit area capacitance between the bottom gate and the nanosheet with $\varepsilon_0$ being the vacuum permittivity, $\varepsilon = 3.9$ the dielectric constant of SiO$_2$, and $d = 300$ nm the thickness of SiO$_2$. In the above relation, $V_{BG}^{th}$ is the threshold voltage at which the conductance $G$ goes to zero. In our case, to extract the threshold, a line fit to the measured $G - V_{BG}$ curve in Fig. 1d is made (see Supplementary Fig. 1a). Then by extending the fitting line to intersect the horizontal axis, we obtain $V_{BG}^{th}$. In this way, we have estimated out a carrier density of $n = 7.2 \times 10^{11}$ cm$^{-2}$ at $V_{BG} = -5$ V and $V_{TG} = 0$ V, at which the measured conductance takes a value of $G \sim 9e^2/h$. Note that along the red contour line in Fig. 1c, the measured conductance stays at the same value of $G \sim 9e^2/h$ and thus the carrier density in the nanosheet stays, to a good approximation, at the same value of $n = 7.2 \times 10^{11}$ cm$^{-2}$. Similarly, the yellow contour line in Fig. 1c displays the measurements at a conductance of $G \sim 5e^2/h$ and a carrier density of $n = 4.3 \times 10^{11}$ cm$^{-2}$ in the nanosheet. The electron mobility in the nanosheet is estimated from $\mu = \sigma/ne$, where $\sigma = \frac{GL}{W}$ is the sheet conductivity with $L$ being the channel length (i.e., the distance between the two inner contact electrodes, 1.1 μm in this device) and $W$ being the channel width (i.e., the width of the nanosheet, 550 nm in this device). Since the conductance is approximately a linear function of $V_{BG}$ and the same is for the electron density in the nanosheet, the same electron mobility of $\mu \sim 6000$ cm$^2 \cdot$V$^{-1} \cdot$s$^{-1}$ in the nanosheet is extracted at both $G \sim 9\ e^2/h$ and $G \sim 5\ e^2/h$. The electron mean free path in the nanosheet can be estimated from $L_e = \frac{\hbar \mu}{e}\sqrt{2\pi n}$, where $\hbar = \frac{h}{2\pi}$ with $h$ being the Planck constant, giving $L_e \sim 84$ nm at $n = 7.2 \times 10^{11}$ cm$^{-2}$ ($G \sim 9\ e^2/h$) and $L_e$~ 65 nm at $n = 4.3 \times 10^{11}$ cm$^{-2}$ ($G$ ~ 5 $e^2/h$). A larger value of $L_e$ obtained at the higher electron density could be due to screening of scattering by electrons in the nanosheet. For comparison, it is worthwhile to note that the Fermi wavelength can be estimated as $\lambda_F = \sqrt{2\pi/n}$~30 nm at the carrier density of $n$ =7.2 × 10$^{11}$ cm$^{-2}$, which is close to the thickness of the nanosheet. Thus, only one or few 2D electron subbands in the InSb nanosheet are occupied and the InSb nanosheet is dominantly a 2D electron system. The same analysis based on the top gate transfer characteristics should give the similar estimations for the carrier density and the mobility at the same setting of $V_{BG}$ and $V_{TG}$. According to this, we have extracted a value of $\varepsilon$~ 6.5 for the dielectric constant of the top gate dielectric HfO2 using the $G - V_{TG}$ curve shown in Fig. 1e and the carrier densities extracted through the $G - V_{BG}$ curve (see Supplementary Note I for detail).

Quantum transport characteristics of the InSb nanosheet

In a quantum diffusive device, the electron transport can be characterized by a set of transport length scales, including phase coherence length ($L_\varphi$), SOI length ($L_{SO}$), and mean free path ($L_e$). In order to determine all these lengths in the InSb nanosheet, we have performed detailed magnetotransport measurements for the dual-gate InSb nanosheet device at low magnetic fields. Figure 2a shows the measured magnetoconductance, $\Delta G = G(B) - G(B = 0)$, at different $V_{BG}$ with top-gate voltage set at $V_{TG} = 0$ V. Here, the magnetic field $B$ is applied perpendicular to the nanosheet. It is seen that the measured magnetoconductance displays a peak in the vicinity of $B = 0$, i.e., the weak antilocalization (WAL) characteristics. The WAL arises from quantum interference in the presence of strong SOI and gives a positive quantum correction to the conductance at zero magnetic field. It is also seen that at $V_{BG} = 0$ V, a well-defined WAL peak is observed, but the peak becomes less pronounced as $V_{BG}$ decreases.

For a 2D diffusive system, the low-field magnetoconductance is well described by the Hikami-Larkin-Nagaoka (HLN) quantum interference theory[43]. Assuming that the electron transport in the InSb nanosheet is in the 2D diffusion regime, the quantum correction to the low-field magnetoconductance is given by

$$\Delta G(B) = -\frac{e^2}{\pi h}[\frac{1}{2}\Psi\left(\frac{B_\varphi}{B}+\frac{1}{2}\right) + \Psi\left(\frac{B_e}{B}+\frac{1}{2}\right)$$
$$-\frac{3}{2}\Psi\left(\frac{(4/3)B_{SO}+B_\varphi}{B}+\frac{1}{2}\right) - \frac{1}{2}\ln\left(\frac{B_\varphi}{B}\right)$$
$$-\ln\left(\frac{B_e}{B}\right) + \frac{3}{2}\ln(\frac{(4/3)B_{SO}+B_\varphi}{B})]. \qquad (1)$$

Here, $\Psi(x)$ is the digamma function. Three subscripts, φ, SO, and e, in the above equation denote inelastic dephasing, spin-orbit scattering, and elastic scattering processes, respectively. $B_{\varphi,SO,e}$ are the characteristic fields for the three scattering mechanisms and are given by $B_{\varphi,SO,e} = \hbar/(4eL^2_{\varphi,SO,e})$. The measured low-field magnetoconductance data at different $V_{BG}$ shown in Fig. 2a are fitted to Eq. (1) using $L_\varphi$, $L_{SO}$, and $L_e$ as fitting parameters (see further detail in Methods). The black solid lines in Fig. 2a are the results of the fits.

Figure 2b shows the extracted $L_\varphi$, $L_{SO}$, and $L_e$ in the InSb nanosheet from the fits at $V_{TG} = 0$ V as a function of $V_{BG}$. As shown in Fig. 2b, $L_\varphi$ is strongly dependent on $V_{BG}$, while $L_{SO}$ and $L_e$ show weak $V_{BG}$ dependences and stay at values of $L_{SO} \sim 130$ nm and $L_e \sim 80$ nm. Here, we note that the extracted $L_e \sim 80$ nm is in good agreement with the values extracted above from the gate transfer characteristics. The weak $V_{BG}$ dependence of $L_e$ arises from the fact that at the low temperature we have considered, $L_e$ is primarily given by the distribution of scattering carriers, such as charged impurities and lattice defects, in the conduction InSb channel and the dielectric $SiO_2$ layer, as well as at the InSb-$SiO_2$ interface, and the distribution of scattering centers should be insensitive to a change in the gate voltage in the range we have considered. The $L_{SO}$ also shows a weak $V_{BG}$ dependence because it primarily depends on the perpendicular electric field penetrated through the InSb nanosheet, which is only weakly dependent on $V_{BG}$ when the InSb nanosheet is at open conduction state. At $V_{BG} = 0$ V (a high carrier density case), the extracted $L_\varphi$ reaches to ~ 530 nm. As $V_{BG}$ sweeps from 0 to −13 V, $L_\varphi$ decreases rapidly to ~180 nm, indicating that the dephasing is stronger at a lower carrier density. The physical origin of this increase in $L_\varphi$ with increasing carrier density is that, at this low temperature, the dephasing arises predominantly from electron-electron interaction with small energy transfers, in the form of electromagnetic field fluctuations generated by the motions of neighboring electrons (the Nyquist dephasing mechanism[44]), and such fluctuations get to be diminished at a higher carrier density and thus an increased bottom gate voltage due to stronger charge screening. It is worthwhile to emphasized that $L_\varphi$ is one order of magnitude larger than the thickness of the nanosheet. This, together with the fact that the typical Fermi wavelength $\lambda_F \sim 30$ nm is close to the thickness of the nanosheet, supports our assumption that the transport in the nanosheet is of a 2D nature. In addition, the extracted $L_e \sim 80$ nm is one order of magnitude smaller than the distance between the two inner contact electrodes, indicating that the transport in the nanosheet is in the diffusion regime.

There are several possible mechanisms responsible for the spin relaxation process in the nanosheet. One is the Elliot-Yafet mechanim[45,46], i.e., the spin randomization due to momentum scattering. In the Elliot-Yafet mechanism, the spin relaxation length can be estimated out as[39,47]

$L_{SO,EY} = \sqrt{\frac{3}{8}} \cdot \frac{E_G}{E_F} \cdot L_e \cdot \frac{(E_G+\Delta_{SO})(3E_G+2\Delta_{SO})}{\Delta_{SO}(2E_G+\Delta_{SO})} \geq$ 500 nm, using the bandgap $E_G$ = 0.23 $e$V, the Fermi energy $E_F = \frac{\hbar^2 \pi n}{m^*} \leq$ 50 m$e$V (with $n \leq 7.2 \times 10^{11}$cm$^{-2}$), bulk spin-orbit gap[48] $\Delta_{SO}$ ~ 0.8 $e$V, and the mean-free path $L_e$ ~ 80 nm. The estimated $L_{SO,EY}$ is much larger than the experimentally extracted value of $L_{SO}$ ~ 130 nm. Therefore, the Elliot-Yafet mechanism does not play a key role in our system. Another one is the D'yakonov-Perel' mechanism[49], which considers the spin precession between scattering events. Since the InSb nanosheets used in our device is a zincblende crystal and the current flow would take along a <111> or a <110> crystallographic direction[27], the Dresselhaus SOI[50] would be either absent or negligible[51]. Based on the above analyses, we expect that the Rashba SOI[52] is the primary cause of spin relaxation in the InSb nanosheet. This expectation is also consistent with our designed device structure with an enhanced structural asymmetry. Hence we can obtain a Rashba spin-orbit strength of $\alpha_R$~0.42 $e$V Å according to $L_{SO} = \hbar^2/m^*\alpha_R$, where $m^* = 0.014\ m_0$ denotes the effective mass of electrons in InSb with $m_0$ being the free electron mass. The spin-orbit energy can be determined as $E_{so} = \frac{m^*\alpha_R^2}{2\hbar^2}$ ~ 160 μ$e$V in the InSb nanosheet. In comparison with most commonly employed III-V narrow bandgap semiconductor nanostructures with a strong SOI, the extracted spin-orbit strength of $\alpha_R$~0.42 $e$V Å in our InSb nanosheet from the low-field magnetotransport measurements shown in Fig. 3 is smaller than but comparable to the values of 0.5-1 $e$V Å found in InSb nanowires[39], but is significantly larger than the values of ~ 0.16 $e$V Å found in InAs nanowires[53]. In addition, our extracted spin-orbit strength in the InSb nanosheet is an order of magnitude larger than the values reported previously for InSb and InAs quantum wells[41,54]. Thus, the extracted $\alpha_R$ ~ 0.42 $e$V Å in our InSb nanosheet corresponds to a strong SOI found in a III-V narrow bandgap semiconductor nanostructure.

Tuning the SOI in the InSb nanosheet by dual-gate voltage

The SOI of the Rashba type is tunable by applying an electric field perpendicularly through the InSb nanosheet. Such an electric field can be achieved and tuned by a voltage $V_D$ applied over the dual gate. For example, with $V_{TG}$ being set at 0 V, we could sweep $V_{BG}$ to gradually change $V_D$ and thus the electric field through the nanosheet. However, as we showed above, sweeping $V_{BG}$ only also tunes the carrier density in the nanosheet. To demonstrate the manipulation of SOI solely via the vertical electric field in the nanosheet, the carrier density in the nanosheet ought to be fixed. In the present work, this is achieved by performing magnetotransport measurements along an equal conductance contour line, in which the carrier density in the nanosheet approximately stays at a constant value, but the dual-gate voltage, $V_D = V_{TG} - V_{BG}$, is tuned continuously. Figure 3a shows magnetoconductance traces measured along a contour line of $G \sim 9e^2/h$ (the red contour line in Fig. 1c) at several values of $V_D$. It is seen that all the measured magnetoconductance traces show the WAL characteristics. To extract the transport length scales as a function of $V_D$, we fit these measured magnetoconductance traces to Eq. (1). The black solid lines in Fig. 3a show the results of the fits. Figure 3b displays the characteristic transport lengths $L_\varphi$, $L_{SO}$, and $L_e$ extracted from the fits. It is shown that $L_\varphi$ stays at a constant value of ~460 nm, independent of $V_D$. This is in good agreement with the fact that $L_\varphi$ is mainly influenced by carrier density and temperature, but not by an electric field applied perpendicular to the nanosheet. The same is also true for $L_e$, which is found to stay at a value of ~ 85 nm. However, $L_{SO}$ shows a strong dependence on $V_D$. As seen in Fig. 3b, $L_{SO}$ is monotonically increased from ~130 to ~390 nm as $V_D$ changes from −2 to 11 V, indicating that the SOI strength becomes weaker as $V_D$ moves towards more positive values. Figure 3c shows the magnetoconductance traces measured along a constant conductance contour line of $G \sim 5e^2/h$ (the yellow contour line in Fig. 1c) at varying $V_D$ from −4.4 to 10.7 V. Again, the WAL characteristics are observed in the measurements. The black solid lines in Fig. 3c show the fits of the measurements to Eq. (1) and Fig. 3d shows the transport lengths extracted from the fits. Again, it is seen that with varying $V_D$, $L_\varphi$ stays at a value of ~340 nm and $L_e$ stays at a value of ~78 nm. i.e., both are independent of $V_D$. However, $L_{SO}$ is seen to increase from ~130 to ~270 nm as $V_D$ is tuned from −4.4 to 10.7 V. Our results presented in Fig. 3 clearly demonstrate that the SOI in the InSb nanosheet of our dual-gate device can be effectively tuned by applying a voltage over the dual gate without a change in the carrier density in the nanosheet. The achieved change in $L_{SO}$ from 130 to 390 nm corresponds to a change in the spin-orbit strength from 0.42 to 0.14 $eV \cdot Å$ and a change in the spin-orbit energy from 160 to 18 $\mu eV$.

We have also performed the dual-gate voltage $V_D$ dependent measurements of the transport characteristics lengths $L_\varphi$, $L_{SO}$, and $L_e$ in the InSb nanosheet along the constant conductance contour lines of ~2.6 and ~$1.1e^2/h$, and an efficient tuning of SOI in the nanosheet by the dual-gate voltage $V_D$ is again observed (See Supplementary Note V). All the results presented in the present section (and in Supplementary Fig. 5) manifest that the SOI in the InSb nanosheet in a dual-gate structure can be efficiently tuned by a voltage applied to the dual gate at largely different but fixed carrier densities of the nanosheet.

Band diagram and intrinsic Rashba SOI in the InSb nanosheet

It is important to emphasize that the experimentally extracted Rashba spin-orbit length $L_{SO}$ is small, indicating a strong SOI, even at $V_D = 0$ V. This seemly unexpected observation however reveals the presence of an intrinsic structural asymmetry even in the absence of a voltage difference between the top and bottom gates due to band offsets appeared in the HfO$_2$ – InSb - SiO$_2$ heterostructure. To show this, the energy band diagram in the vertical direction is simulated using commercially available software COMSOL. The simulation is mainly based on Possion's equations and takes the material parameters of HfO$_2$, InSb and SiO$_2$, including bandgaps, dielectric constants, electron effective masses, and electron affinities, as inputs (see Supplementary Table I for material parameters). Figure 4a displays the simulated energy band diagram of the HfO$_2$-InSb-SiO$_2$ structure at $V_D = 0$ V (with $V_{TG} = V_{BG} = -0.33$V) and carrier density $n = 7.2 \times 10^{11}$ cm$^{-2}$ in the InSb nanolayer. The conduction band and the valence band edges exhibit a noticeable bending even at $V_D = 0$ V, illustrating the presence of an intrinsic structure asymmetry in the InSb nanosheet. Figure 4b shows a zoom-in view of the simulated conduction band edge in the InSb nanosheet at three values of $V_D$. The green, blue, and red solid lines represent the conduction band edges at $V_D = 0$, $-2$ (with $V_{TG} = -0.46$ V and $V_{BG} = 1.54$ V), and 11 V ($V_{TG} = 0.4$ V and $V_{BG} = -10.6$ V), respectively, and the carrier density of $n = 7.2 \times 10^{11}$ cm$^{-2}$ in the InSb nanosheet. With pushing the dual-gate voltage from $V_D = 0$ V to $V_D = -2$ V, we can see that the band bending becomes stronger, indicating an enhanced structural asymmetry and thus a stronger Rashba SOI. On the contrary, by moving from $V_D = 0$ V to $V_D = 11$ V, we compensate the band bending towards a nearly flat band, leading to a reduced asymmetry in the structure and a weaker Rashba SOI. These simulated results are fully in line with the experimental observations. Based on the simulations, the strength of the vertical, effective mean electric field in the InSb nanosheet can be estimated. It is found that the field strength gradually increases when changing from $V_D = 11$ V to $V_D = -2$ V (see Supplementary Fig. 3), providing a numerical evidence for tunable structural asymmetry and thus tunable SOI in the nanosheet. Furthermore, the material-specific, Fermi-level-dependent Rashba prefactor $r_R$, defined via the Rashba coefficient $\alpha_R = r_R E$ with $E$ being the strength of the perpendicular electric field in the InSb nanosheet[48], can be determined (see Supplementary Note IV for detail). The Rashba prefactors extracted for the InSb nanosheet are $r_R = 4.26$ $e \cdot$nm$^2$ at carrier density $n = 7.2 \times 10^{11}$ cm$^{-2}$ and $r_R = 3.48$ $e \cdot$nm$^2$ at carrier density $n = 4.3 \times 10^{11}$ cm$^{-2}$.

Temperature effects

Figure 5a shows the measured low-field magnetoconductance of the device at $V_{\text{BG}} = 1.54$ V and $V_{\text{TG}} = -0.46$ V at temperatures of 1.9 to 20 K. At temperature $T = 1.9$ K, a sharp WAL peak is seen in the vicinity of zero magnetic field. As the temperature increases, both the height of the WAL peak and the fluctuation magnitude of the UCF patterns become gradually suppressed, although they still remain visible at $T = 20$ K. Again, we fit these measured magnetoconductance data to Eq. (1) and plot the results in black solid lines in Fig. 5a. Extracted $L_\varphi$, $L_{\text{SO}}$ and $L_e$ from the fits are displayed in Fig. 5b. It can be found that both $L_{\text{SO}}$ and $L_e$ are weakly dependent on temperature, while $L_\varphi$ shows a strong temperature dependence, decreasing rapidly from ~470 nm to ~210 nm with increasing temperature from 1.9 to 20 K. The temperature dependence of $L_\varphi$ is found to follow a power law of $L_\varphi \sim T^{-0.38}$ (see the solid line in Fig. 5b). The power index of −0.38 falls between the values of −1/2 and −1/3, corresponding to the Nyquist dephasing processes[44] in a 2D ($T^{-1/2}$) and a 1D system ($T^{-1/3}$). Since $L_\varphi$ is in the same order of length scale as the width of the conduction channel (~ 550 nm), the transport in the InSb nanosheet is more likely in an intermediate regime between the 1D and 2D limits[55].

In summary, a dual-gate planar device made from a single-crystalline zincblende InSb nanosheet is fabricated and the quantum transport properties of the InSb nanosheet in the device are studied by low-field magnetotransport measurements. Carrier density, mean free path, the coherence length, and SOI strength in the InSb nanosheet are extracted. It is shown that the measured low-field magnetoconductance can be excellently described by the 2D diffusive HLN quantum transport theory and exhibits the WAL characteristics. The origin of the WAL characteristics is identified as the presence of strong SOI of the Rashba type in the InSb nanosheet. By performing the magnetoconductance measurements of the InSb nanoshhet at constant carrier densities, we demonstrate that the Rashba SOI strength can be efficiently tuned by a voltage applied over the duel gate. We also observe the presence of a strong SOI in the InSb nanosheet at zero dual-gate voltage. By simulations for the band diagrams of the device structure, we identify the origin of this intrinsic SOI in the InSb nanosheet as the presence of band bending in the nanosheet even at zero dual-gate voltage. The strong and tunable Rashba SOI in the InSb nanosheet, demonstrated in this work, lays the groundwork for employing this emerging layered material in the developments of spintronics, spin qubits, and topological quantum devices.

**Methods**

Material growth

High-quality, free-standing, single-crystalline, pure zincblende phase InSb nanosheets used in this work are grown by molecular-beam epitaxy (MBE) on top of InAs nanowires on a Si (111) substrate. The growth process starts by depositing a thin layer of Ag on the Si substrate in an MBE chamber. The film is subsequently annealed *in situ* to form Ag nanoparticles. Thin InAs nanowires are then grown with these Ag nanoparticles as seeds. The InSb nanosheets are grown on top of the InAs nanowires by abruptly switching the group-V source from As to Sb and with an increased Sb flux. High-resolution transmission electron microscopy and scanning electron microscopy analyses show that the as-grown InSb nanosheets are of high-quality, pure zincblende phase, single crystals and are up to several micrometers in sizes and down to ~10 nm in thickness. For further detail about the growth process and structural properties of our MBE-grown InSb nanosheets, we refer to Ref. 27.

Device fabrication

For device fabrications, the MBE-grown InSb nanosheets are mechanically transferred from the growth substrate onto an *n*-doped Si substrate covered with a 300-nm-thick layer of $SiO_2$ on top. The Si and $SiO_2$ layers are later used as a global bottom gate and its dielectric. After transferring, contact electrodes are fabricated on selected nanosheets with a thickness of $t\sim30$ nm via a combined process of electron-beam lithography (EBL), electron-beam evaporation (EBE) of a Ti/Au (5/90 nm in thickness) metal bilayer, and lift off. We note that before the metal evaporation, the exposed areas on the InSb nanosheets are chemically etched in a de-ionized water-diluted $(NH_4)_2S_x$ solution to remove the surface oxide and to subsequently passivate the fresh surface. After the contact electrode fabrication, a 20-nm-thick $HfO_2$ dielectric layer is deposited on the sample by atomic layer deposition (ALD). Finally, a Ti/Au (5/90 nm in thickness) metal bilayer top gate is fabricated on each device by the combined process of EBL, EBE and lift off, again. Figure 1a shows a false-colored SEM image of a fabricated device measured for this work and the measurement circuit setup. In this device, four parallel contact electrodes are made on the InSb nanosheet and the distance between the two inner contact electrodes is ~1.1 μm. The top gate covers the entire InSb nanosheet as seen in Fig. 1a and as indicated in the schematic shown in Fig. 1b.

Gate transfer characteristics and magnetotransport measurements

Low-temperature transport measurements of the fabricated devices are carried out in a physical property measurement system (PPMS) cryostat equipped with a uniaxial magnet. The InSb nanosheet conductance is measured in a four-probe configuration to eliminate the impact of the contact resistances using a lock-in technique, in which a 17 Hz AC excitation current $I$ of 100 nA is supplied between the two outer electrodes and the voltage drop $V$ between the two inner contact electrodes is recorded. The nanosheet channel conductance $G$ is obtained numerically from $G = I/V$. For this work, the results of the measurements obtained from a representative device as shown in Fig. 1a are presented. The measurements are carried out with magnetic fields applied perpendicular to the nanosheet plane at temperatures of $T$ = 1.9 to 20 K.

Fitting of the measured magnetoconductance data to the HLN formula

To extract the characteristic transport lengths of $L_\varphi$, $L_{SO}$, and $L_e$ in the InSb nanosheet, the measured data are fitted to Eq. (1), based on the least square method, using both the "curve_fit" function in the SciPy package written in Python and the non-linear fit program in the Origin software for crosscheck. The two fitting procedures give almost the same results. The fitting bounds are set in order to make the corresponding length scales vary in a reasonable range. For example, we set the fitting bound of $L_e$ as $L_e \leq 200$ nm. The range of magnetic fields $B$ is chosen to be $|B| \leq 20$ mT in all the fittings presented in this work in order to make the low field condition of Eq. (1) satisfied.

Band diagram simulation

To simulate the energy band diagrams of the $HfO_2$ - InSb - $SiO_2$ heterostructure in the device, Poisson's equations are solved using commercially available program COMSOL in compliance with the boundary conditions of the system. An effective one dimensional model with three sections representing three different materials, $HfO_2$, InSb and $SiO_2$, is considered. Material properties employed in the simulation, include bandgaps, dielectric constants, electron effective masses, and electron affinities, are listed in Supplementary Table I. The carrier density in the InSb nanosheet and the boundary conditions used in the simulation are acquired from the experiments. We first show the different degrees of band bending, i.e., the different degrees of asymmetry, when various voltages are applied to the top and bottom gates (Fig. 4 and Supplementary Fig. 3a).

The carrier density distribution inside the InSb layer can also be calculated (see Supplementary Fig. 3c). It is seen that the carrier density is non-uniformly distributed, consistent with the conduction band bending profile obtained. The quantitative analysis of the asymmetry is carried out from the calculated effective electric field strength inside the InSb layer shown in Supplementary Fig. 3b.

## Data Availability

The data supporting the findings of this study are available within the article and its Supplementary Information file. Additional data including simulation codes are available from the corresponding author upon reasonable request.

## Acknowledgements

This work is supported by the Ministry of Science and Technology of China through the National Key Research and Development Program of China (Grant Nos. 2017YFA0303304, 2016YFA0300601, 2017YFA0204901, and 2016YFA0300802), the National Natural Science Foundation of China (Grant Nos. 11874071, 91221202, 91421303, 11274021, and 61974138), and the Beijing Academy of Quantum Information Sciences (Grant No. Y18G22). D.P. also acknowledges the support from the Youth Innovation Promotion Association of the Chinese Academy of Sciences (No. 2017156).

## Author contributions

H.Q.X. conceived and supervised the project. Y.C. and S.H. fabricated the devices and carried out the transport measurements. D.P. and J.H.Z. grew the materials. J.X. and L.Z. participated in the device fabrication and measurements. Y.C. performed the band diagram simulations. Y.C., S.H. and H.Q.X. analyzed the data and wrote the manuscript with contributions from all authors. All authors contributed to the discussion of the results and the interpretation of the data acquired.

## Competing interests

The authors declare no competing interests.

**Figure Captions**

**Fig. 1 Dual-gate InSb nanosheet device and its gate transfer characteristics. a**, False-colored SEM image of the device studied in this work and measurement circuit setup. The device is fabricated from an MBE-grown InSb nanosheet (transport channel) on an $n$-doped Si substrate (bottom gate) covered by a layer of SiO$_2$ on top (bottom gate dielectric). The InSb nanosheet is contacted by four stripes of Ti/Au (contact electrodes) and is then covered by depositing a layer of HfO$_2$ (top gate dielectric) and a metal bilayer of Ti/Au (top gate). The nanosheet has a width of ~550 nm and a thickness of ~30 nm. The separation between the two inner electrodes is ~1.1 μm. **b**, Schematic view of the layer structure of the device. **c**, Conductance $G$ measured for the device as functions of top-gate voltage $V_{TG}$ and bottom-gate voltage $V_{BG}$ (gate transfer characteristics) at a temperature of $T = 1.9$ K. In the measurements, a 17-Hz AC current ($I$) of 100 nA is applied through the two outer electrodes, and the voltage drop ($V$) between the two inner electrodes is recorded and is then converted to the conductance through $G = I/V$. The red and yellow solid lines denote the constant conductance contours of ~9 and ~5$e^2/h$, respectively. **d**, $G$ as a function of bottom-gate voltage $V_{BG}$ at $V_{TG} = 0$ V. **e**, $G$ as a function of top-gate voltage $V_{TG}$ at $V_{BG} = 0$ V.

**Fig. 2 Magnetotransport measurements at various bottom gate voltages. a**, Low-field magnetoconductance, $\Delta G = G(B) - G(B = 0)$, measured for the device shown in Fig.1a at various bottom-gate voltages $V_{BG}$ at temperature $T = 1.9$ K. The bottom trace shows the measured magnetoconductance data at $V_{BG} = 0$ V and all other measured magnetoconductance traces are successively vertically offset for clarity. Here the top-gate voltage is set at $V_{TG} = 0$ V. The black solid lines are the theoretical fits of the experimental data to the HLN equation [Eq. (1)]. **b**, Phase coherence length $L_\varphi$, spin-orbit length $L_{SO}$, and mean free path $L_e$ in the InSb nanosheet extracted from the fits as a function of $V_{BG}$.

**Fig. 3 Magnetotransport measurements at various voltages applied over the dual gate at constant carrier densities in the InSb nanosheet. a**, Low-field magnetoconductance $\Delta G$ measured for the device at a constant conductance of ~ 9 $e^2/h$ and temperature $T = 1.9$ K at various values of dual-gate voltage $V_D = V_{TG} - V_{BG}$. The bottom trace shows the measured magnetoconductance data at $V_D = -2$ V and all other measured magnetoconductance traces are successively vertically offset for clarity. The black solid lines are the theoretical fits of the

experimental data to the HLN equation [Eq. (1)]. **b**, Phase coherence length $L_\varphi$, spin-orbit length $L_{SO}$, and mean free path $L_e$ extracted from the fits in **a** as a function of $V_D$. **c**, The same as **a** but measured for the device at a constant conductance of ~5 $e^2/h$. Here, The bottom trace shows the measured magnetoconductance data at $V_D = -4.4$ V and all other measured magnetoconductance traces are again successively vertically offset for clarity. **d**, Phase coherence length $L_\varphi$, spin-orbit length $L_{SO}$, and mean free path $L_e$ extracted from the fits in **c** as a function of $V_D$.

**Fig. 4 Simulated band diagrams. a**, Simulated energy band diagram of the HfO$_2$-InSb-SiO$_2$ structure in the device at $V_D = 0$ V (with $V_{BG} = V_{TG} = -0.33$ V). The conduction and valence band edges $E_c$ and $E_v$, are marked by the dark blue and light blue solid lines, while the Fermi level $E_F$ is marked by the dark blue dashed line. **b**, Zoom-in plots (red, blue and green solid lines) showing the details of the conduction band edges inside the InSb nanosheet at three different values of $V_D$. The red, blue and green dashed lines mark the Fermi levels obtained in the simulations at the three different values of $V_D$.

**Fig. 5 Temperature-dependent magnetotransport measurements. a**, Low-field magnetoconductance $\Delta G$ measured for the device at $V_{BG} = 1.54$ V and $V_{TG} = -0.46$ V, corresponding to a conductance value of ~$9e^2/h$ and a dual-gate voltage value of $V_D = -2$ V, at different temperatures $T$. The bottom trace shows the measured magnetoconductance data at $T = 1.9$ K and all other measured magnetoconductance traces are successively vertically offset for clarity. **b**, Phase coherence length $L_\varphi$, spin-orbit length $L_{SO}$, and mean free path $L_e$ extracted by fitting the measured data to Eq. (1) as a function of temperature $T$. The red solid line is a power-law fit to the extracted phase coherence length $L_\varphi$, showing $L_\varphi \sim T^{-0.38}$.

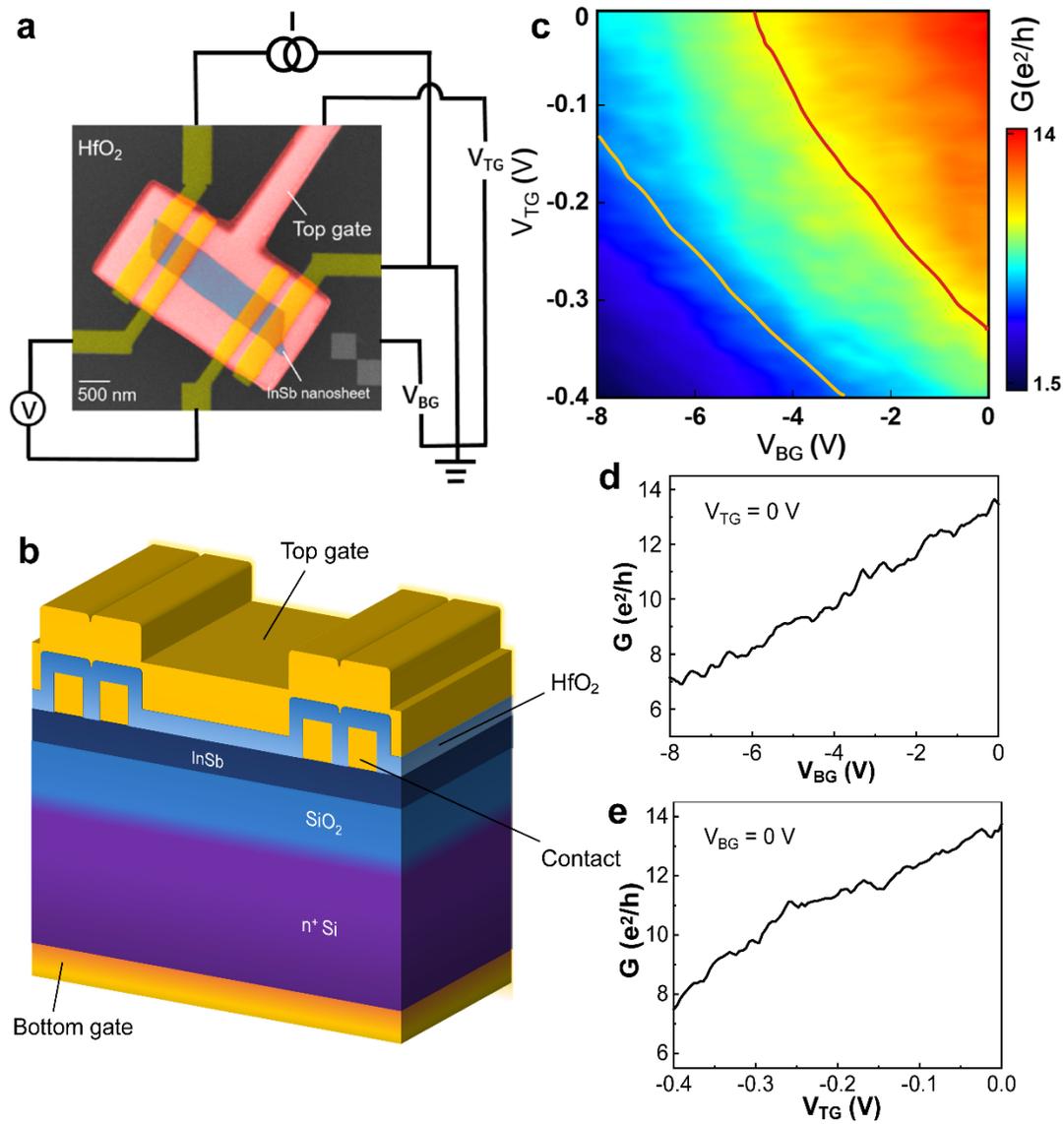

Figure 1, Yuanjie Chen *et al*.

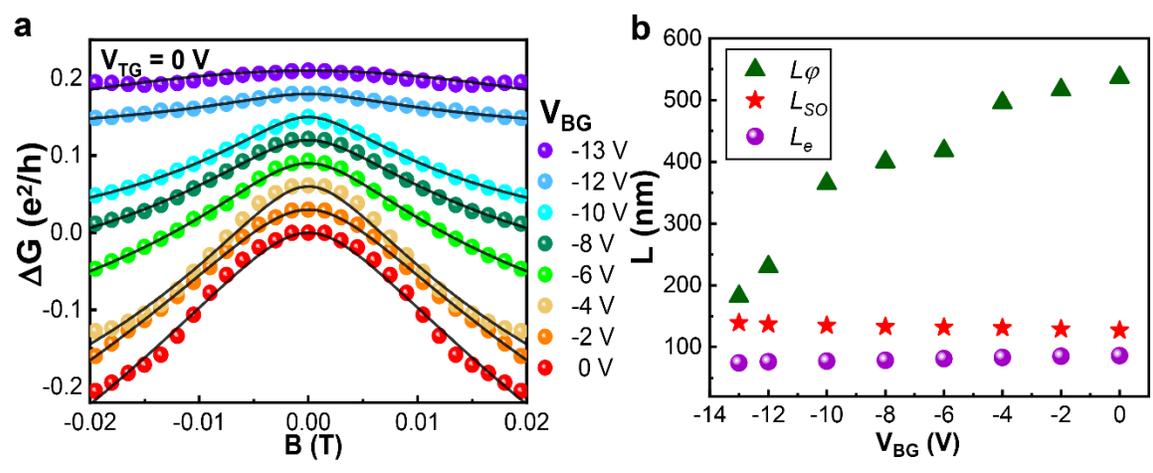

Figure 2, Yuanjie Chen *et al*.

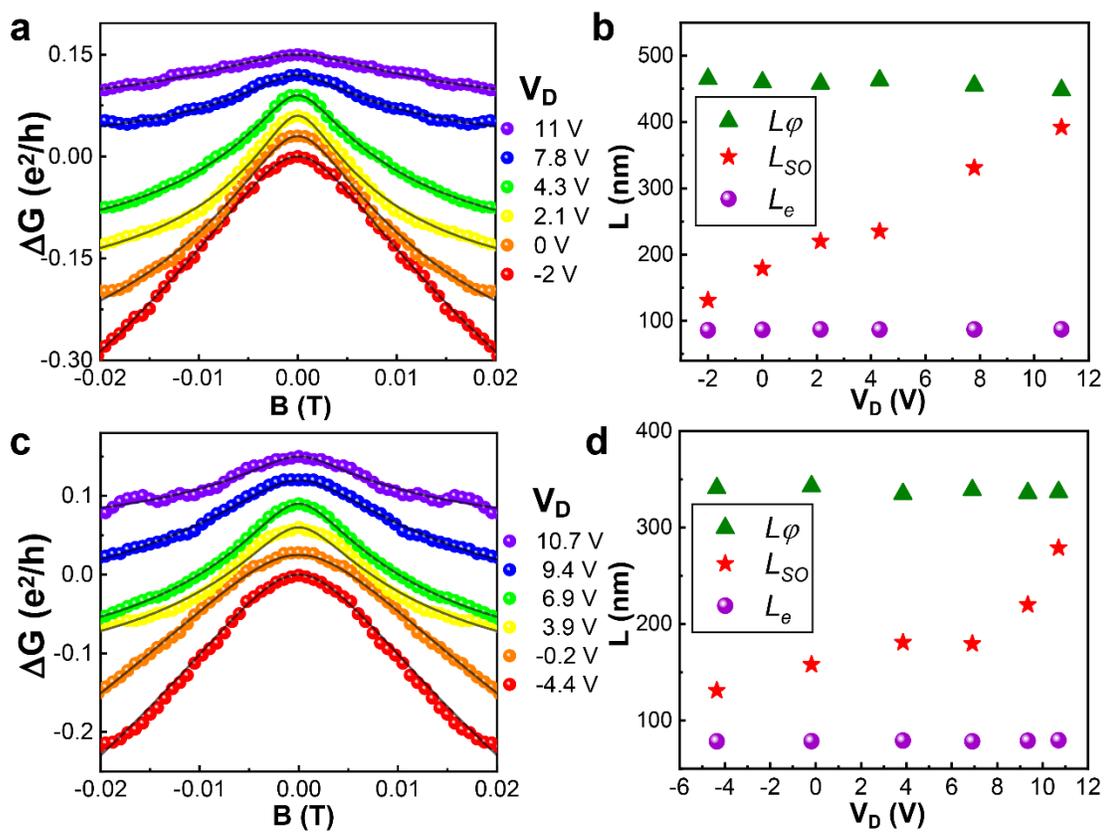

Figure 3, Yuanjie Chen *et al*.

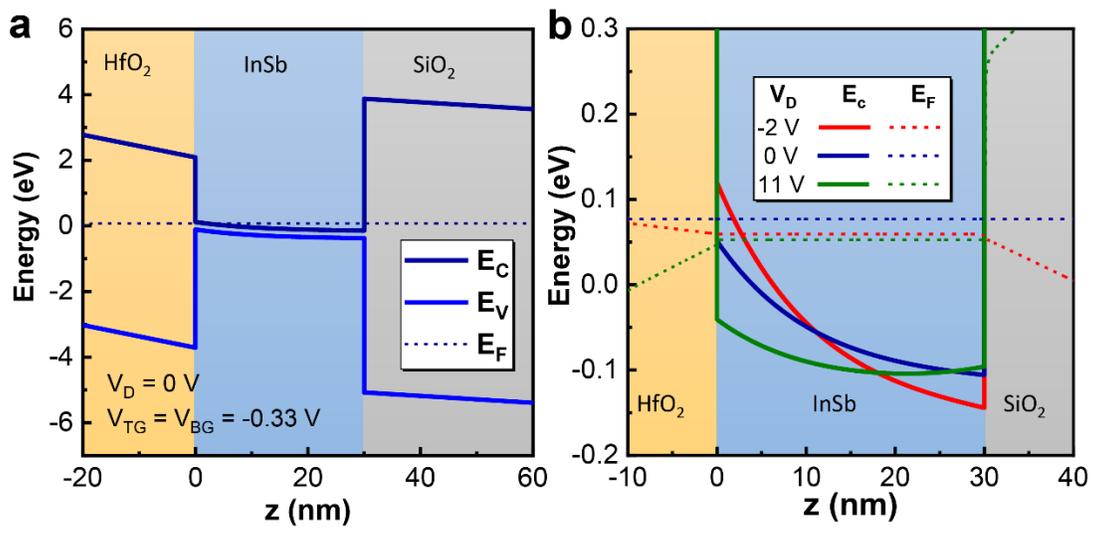

Figure 4, Yuanjie Chen *et al*.

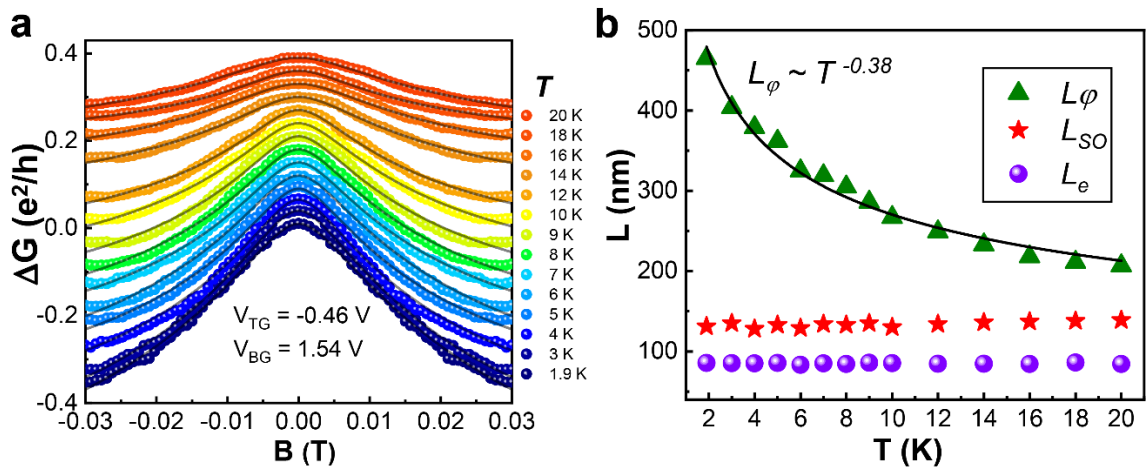

Figure 5, Yuanjie Chen *et al*.

# Supplementary Information for
# Strong and tunable spin-orbit interaction in a single crystalline InSb nanosheet

Yuanjie Chen[1], Shaoyun Huang[1], Dong Pan[2,3], Jianhong Xue[1], Li Zhang[1], Jianhua Zhao[2,3,*], and H. Q. Xu[1,3,†]

[1]*Beijing Key Laboratory of Quantum Devices, Key Laboratory for the Physics and Chemistry of Nanodevices and Department of Electronics, Peking University, Beijing 100871, China*

[2]*State Key Laboratory of Superlattices and Microstructures, Institute of Semiconductors, Chinese Academy of Sciences, P.O. Box 912, Beijing 100083, China*

[3]*Beijing Academy of Quantum Information Sciences, Beijing 100193, China*

Corresponding authors. Emails: †hqxu@pku.edu.cn; *jhzhao@semi.ac.cn

(Dated: November 16, 2020)

**Supplementary Note I. Determination of the carrier density and carrier mobility in the InSb nanosheet**

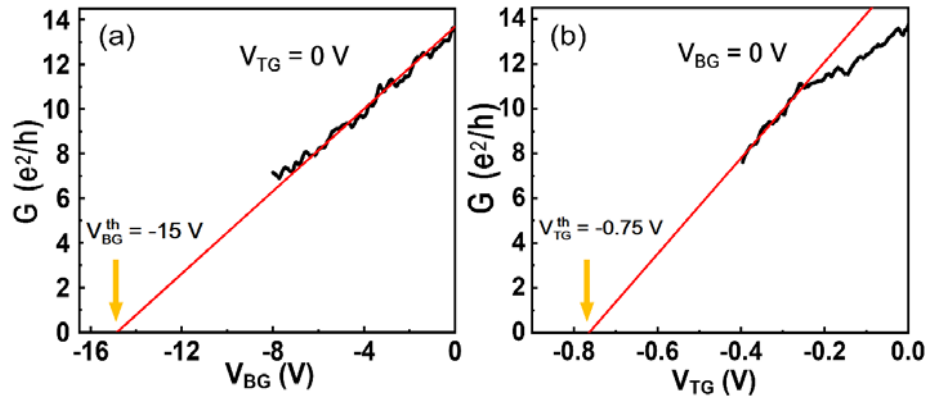

Figure 1. (a) Conductance $G$ (black line) measured for the InSb nanosheet in the dual-gate device as a function of back gate voltage $V_{BG}$ at top gate voltage $V_{TG} = 0$ V and at $T = 1.9$ K. The red line is a linear fit to the measurements. This fitting line is extended to intersect the back-gate axis at which the back gate threshold voltage $V_{BG}^{th} = -15$ V (as marked by a yellow arrow) is extracted. (b) Conductance $G$ (black line) measured for the InSb nanosheet in the dual-gate device as a function of top gate voltage $V_{TG}$ at bottom gate voltage $V_{BG} = 0$ V and at $T = 1.9$ K. The red line is a linear fit to the measurements. This fitting line is extended to intersect the top-gate axis at which the top gate threshold voltage $V_{TG}^{th} = -0.75$ V (as marked by a yellow arrow) is extracted.

In this Supplementary Note, we describe how the carrier density and carrier mobility in the InSb nanosheet of the dual-gate device are extracted from the measured gate transfer characteristics of the device. Figure 1(a) shows the conductance of the nanosheet measured at $T = 1.9$ K as a function of the back gate voltage $V_{BG}$ (back gate transfer characteristics) with top gate voltage fixed at $V_{TG} = 0$ V. The measurements are carried out in a four-probe configuration (see Fig. 1a in the main article for the measurements circuit setup), in which a 17 Hz AC excitation current $I$ of 100 nA is applied through the two outer electrodes and the voltage drop $V$ between the two inner electrodes is reordered. Because the effect of the contact resistances has been eliminated in such four-probe measurements, the conductance of the nanosheet can be obtained directly from $G = I/V$. The carrier density in the nanosheet can be estimated from $n = C_{gs} \times \frac{V_{BG} - V_{BG}^{th}}{e}$, where $C_{gs}$ is the unit area capacitance of the back gate and $V_{BG}^{th}$ is the back gate threshold voltage at which the conductance of the nanosheet goes to zero. Here we estimate $C_{gs}$ using a parallel capacitor model $C_{gs} = \frac{\varepsilon \varepsilon_0}{d}$, where $\varepsilon_0$ is the vacuum permittivity, $\varepsilon$ and $d$ are the dielectric constant and thickness of the gate dielectric. Using $\varepsilon = 3.9$ and $d = 300$ nm for the dielectric layer of SiO$_2$ in this work, we obtain $C_{gs} = 1.15 \times 10^{-8}$ F·cm$^{-2}$. The back gate threshold voltage $V_{BG}^{th}$ is extracted from the measured back-gate transfer characteristics shown in Fig. 1(a) as follows. First, we fit the measured back-gate transfer characteristics by a line (red line in the figure). Then, we extend the fitting line to intersect the back-gate voltage axis and the back gate value at the intercept is the extracted value for $V_{BG}^{th}$. From the measurements shown in Fig. 1(a), $V_{BG}^{th} \sim -15$ V is obtained. The carrier density in the nanosheet can now be evaluated at a given value of $V_{BG}$. For example, at $V_{BG} = -5$ V (corresponding to a case with the nanosheet channel conductance of $G \sim 9\, e^2/h$), a carrier density of $n = 7.2 \times 10^{11}$ cm$^{-2}$ in the nanosheet is obtained, while at $V_{BG} = -9$ V (corresponding to a case with the nanosheet channel conductance of $G \sim 5\, e^2/h$), a carrier density of $n = 4.3 \times 10^{11}$ cm$^{-2}$ in the nanosheet is obtained.

The carrier mobility in the InSb nanosheet can be obtained from $\mu = \sigma/ne$, where $\sigma = \frac{GL}{W}$ is the nanosheet channel conductivity with $L$ being the channel length (i.e., the distance between the two inner contacts, about 1.1 μm in this device) and $W$ the channel width (i.e., the width of the nanosheet, about 550 nm in this device). Here, we note that since both the nanosheet conductance and the carrier density in the nanosheet depend linearly on $V_{BG}$, the extracted carrier mobility from the transfer characteristic measurements (which is often called the field effect mobility) will be independent of $V_{BG}$. Thus, we can evaluate the carrier mobility $\mu$ by setting the back gate voltage value at, e.g., $V_{BG} = -5$ V, at which the carrier density is $n = 7.2 \times 10^{11}$ cm$^{-2}$ and the nanosheet conductance is $G \sim 9\, e^2/h$. The obtained carrier mobility is then $\mu \sim 6000$ cm$^2 \cdot$ V$^{-1} \cdot$ s$^{-1}$. The carrier mean free path in the nanosheet is given by $L_e = \frac{\hbar \mu}{e}\sqrt{2\pi n}$, where $\hbar = \frac{h}{2\pi}$ with $h$ being the Planck constant. From the measured back gate transfer characteristics shown in Fig. 1(a), we obtain $L_e \sim 84$ nm at $n = 7.2 \times 10^{11}$ cm$^{-2}$ (and $G \sim 9e^2/h$) and $L_e \sim 65$ nm at $n = 4.3 \times 10^{11}$ cm$^{-2}$ (and $G \sim 5e^2/h$).

Using the top-gate transfer characteristics shown in Fig. 1(e), similar results for the carrier density and electron mobility in the InSb nanosheet should be extracted. However, since the dielectric constant of HfO$_2$ in our device is an unknown parameter, which has been given to over a wide range of values in the literature, a direct estimation of the carrier density and electron mobility in the InSb nanosheet from the top-gate transfer characteristics is not possible. Nevertheless, using the results obtained above, we can determine the dielectric constant of HfO$_2$ employed in our device. The equation to be used for extraction of the carrier density based on the top-gate transfer characteristics becomes $n = C_{gs} \times \frac{V_{TG} - V_{TG}^{th}}{e}$, where $C_{gs} = \frac{\varepsilon \varepsilon_0}{d}$ with $\varepsilon$ being the unknown dielectric constant of HfO$_2$, $d = 20$ nm the thickness of HfO$_2$, and $V_{TG}^{th}$ the top-gate threshold voltage at which the conductance $G$ goes to zero. Similarly as in Fig. 1(a), $V_{TG}^{th}$ can be extracted from the measurements shown in Fig. 1(b) by a linear fit to the low top-gate voltage data and by extending the fitting line to intersect the top-gate voltage axis. As seen in Fig. 1(b), a result of $V_{TG}^{th} \sim -0.75$ V is obtained. To determine the dielectric constant $\varepsilon$ of HfO$_2$ in our device, we consider the case of the conductance $G \sim 9 e^2/h$ at $V_{TG} = -0.35$ V, corresponding to the case of carrier density $n = 7.2 \times 10^{11}$ cm$^{-2}$ as estimated through the bottom-gate transfer characteristics. By taking this value into the above equation, a value of $\varepsilon \sim 6.5$ can be obtained for the dielectric constant of HfO$_2$ in our device. This value indicates that the HfO$_2$ layer in our device is in good amorphous phase, consistent with the fact that it was grown at a low temperature by atomic layer deposition.

**Supplementary Note II. Comparison between the results obtained by analyses of magnetotransport measurements of the InSb nanosheet using the HLN and ILP theories**

In the main article, the HLN model is utilized in the analysis of our magnetotransport data. This is suitable for a weak disordered system such as InSb nanosheet and other emerging 2D materials, where the electron elastic scattering length, or the mean free path, $L_\text{e}$ is shorter than all other characteristic transport length scales, such as phase coherence length $L_\varphi$ and spin-orbit length $L_\text{SO}$. However, in a clean 2D electron system with a ultrahigh mobility made from a semiconductor heterostructured quantum well, $L_\text{e}$ can be exceedingly longer than $L_\text{SO}$. In this case, the HLN model may no longer be applicable and one might need to invoke the so-called ILP model, developed by Iordanskii, Lyanda-Geller and Pikus[1], in analyses of the magnetotransport measurement data. Here, it is worthwhile to check whether the ILP model can be applied to the magnetotransport data obtained in our device. In the ILP model, the quantum conductance correction to the low-field magnetoconductance is given by

$$\Delta\sigma_\text{ILP} = -\frac{e^2}{4\pi^2\hbar}\left\{\frac{1}{a_0} + \frac{2a_0+1+\frac{H_\text{SO}}{B}}{a_1\left[a_0+\frac{H_\text{SO}}{B}\right]-2\frac{H'_\text{SO}}{B}} + 2\ln\frac{H_\text{tr}}{B} + \psi\left(\frac{1}{2}+\frac{H_\varphi}{B}\right) + 3C - \sum_{n=1}^{\infty}\left[\frac{3}{n} - \frac{3a_n^2 + 2a_n\frac{H_\text{SO}}{B} - 1 - 2(2n+1)\frac{H'_\text{SO}}{B}}{\left[a_n+\frac{H_\text{SO}}{B}\right]a_{n-1}a_{n+1} - 2\frac{H'_\text{SO}}{B}[(2n+1)a_n-1]}\right]\right\}. \quad (1)$$

Here, $H_\text{SO} = H'_\text{SO} + H^3_\text{SO}$ with $H'_\text{SO}$ being Rashba term and $H^3_\text{SO} > 0$ the cubic Dresselhaus term, $a_n = n + \frac{1}{2} + \frac{H_\varphi}{B} + \frac{H_\text{SO}}{B}$, $C$ is Euler's constant, $H_\text{tr} = \frac{\hbar}{4eL_\text{e}^2}$, and $H_\varphi = \frac{\hbar}{4eL_\varphi^2}$. In the calculations using the above equation, the summation of first 10000 terms in the series has been performed and it is checked that a desired convergence has been achieved.

Figure 2 given below shows a comparison between the results of analyses using the HLN and ILP theories, where the magnetoconductance data (dots) measured for our InSb nanosheet device at $T$ = 1.9 K, $V_\text{TG}$ = −0.46 V and $V_\text{BG}$ = 1.54 V, and the best fits to the data by both the HLN (light blue line) and ILP (light green line) models are presented. Clearly, the HLN model yields a satisfactory fit to the measurement data, giving the extracted length scales of $L_\varphi$= 472 nm, $L_\text{SO}$=137 nm and $L_\text{e}$= 88 nm. However, large deviations from the measurements are found in the best fit to the ILP model. In addition, the best fit by the ILP model gives $H'_\text{SO}$= 4.35× $10^{-2}$ T, $H_\varphi$=

$4.04\times 10^{-4}$ T, $H_{\text{tr}}= 2.98\times 10^{-2}$ T, and $H_{\text{so}}^3= 6.2\times 10^{-21}$ T ($\approx 0$ T), and thus the length scales of $L_\varphi= 638$ nm, $L_{\text{SO}}= 61$ nm and $L_{\text{e}} = 74$ nm. Here, both the values of $L_\varphi$ and $L_{\text{e}}$ may look reasonable, but the value of $L_{\text{so}}(< L_{\text{e}})$ looks unphysical for the InSb nanosheet in our device. Thus, for our InSb nanosheet device, it is more appropriate to use the HLN model, instead of the ILP model, for analyses of our measured magnetoconductance data.

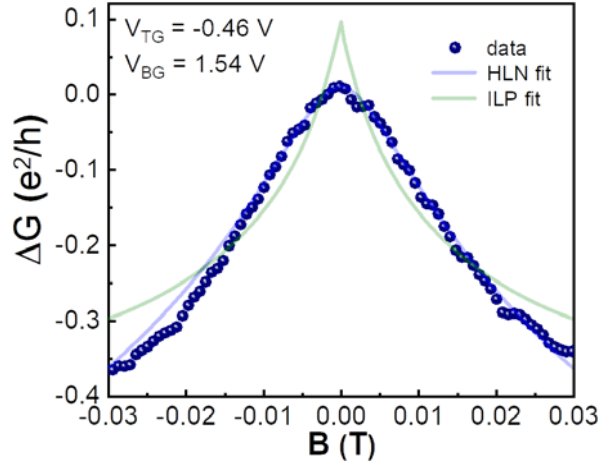

Figure 2. Comparison between the best fits of the magnetoconductance data measured for our InSb nanosheet device at $T = 1.9$ K, $V_{\text{TG}} = -0.46$ V and $V_{\text{BG}} = 1.54$ V to the HLN and ILP models.

**Supplementary Note III. Materials parameters and simulations for energy band diagrams**

To simulate the energy band diagrams of the $HfO_2$-InSb-$SiO_2$ structure in the dual-gate InSb nanosheet device, Poisson's equation is solved using commercially available program COMSOL in compliance with the boundary conditions set in our experiment. Here we assume that each material layer is an infinite two-dimensional structure and we thus need to solve effectively only a one-dimensional Poisson's equation. Material parameters of InSb, $SiO_2$ and $HfO_2$ utilized in the simulations for the energy band diagrams are given in Table I.

Poisson's equation used here to describe the electrostatics of the $HfO_2$-InSb-$SiO_2$ heterostructure has a form of

$$\nabla \cdot (-\epsilon_{\text{r}}\nabla V) = q(p - n + N_{\text{d}}^+ - N_{\text{a}}^-), \qquad (2)$$

where $V$ is the electric potential, $\epsilon_{\text{r}}$ is the dieletric constant of the material, $q$ is the elementary charge, $p$ and $n$ are the hole and electron densities, $N_d^+$ and $N_a^-$ are the ionized donor and acceptor concentrations, respectively. The energies of the conduction and valence band edges can be calculated as, $E_c = -(eV + \chi_0)$ and $E_V = -(eV + \chi_0 + E_{g,0})$, where $\chi_0$ and $E_{g,0}$ are the electron affinity and bandgap of a material. The continuity conditions at the interface of two different materials are, $\boldsymbol{n} \cdot (D_1 - D_2) = 0$ and $E_{F,1} = E_{F,2}$, where $\boldsymbol{n}$ denotes the normal vector of the interface, $D_{1,2}$ and $E_{F,1,2}$ are the electric displacements and electron Fermi levels in the two materials.

Table I. Material parameters of InSb, $SiO_2$ and $HfO_2$ utilized in the simulations for the $HfO_2$-InSb-$SiO_2$ heterostructure in the dual-gate InSb nanosheet device at $T = 2$ K.

| Material | Bandgap (eV) | Dielectric Constant | Electron Effective Mass ($m_0$) | Electron Affinity (eV) |
|---|---|---|---|---|
| InSb | 0.23[1] | 16.8 | 0.014 | 4.77[2] |
| $HfO_2$ | 5.8 | 6.5[3] | 0.11 | 2.8[4] |
| $SiO_2$ | 8.95 | 3.9 | 0.3 | 0.75[5] |

[1] Littler, C. L. & Seiler, D. G. *Appl. Phys. Lett.* **46**, 986 (1985).
[2] Freeouf, J. L., and J. M. Woodall. *Appl. Phys. Lett.* **39**, 727-729 (1981).
[3] From this work.
[4] Sayan, S., Eric Garfunkel, and S. Suzer. *Appl. Phys. Lett.* **80**, 2135-2137 (2002).
[5] Fujimura, Nobuyuki, et al. *Japanese Journal of Applied Physics* **55**. 08PC06 (2016)

Figure 3(a) is the zoom-in figure of Fig. 4b in the main article which displays the calculated profiles of the conduction band edge inside the InSb layer at the experimental condition of the carrier density $n = 7.2 \times 10^{11}$ cm$^{-2}$ and the conductance $G \sim 9e^2/h$ with three different values of voltage $V_D$ applied over the dual gate. Clearly the conduction band edge inside the InSb layer is bended, leading to the presence of an electric field in the layer. Figure 3(b) shows the calculated electric field distribution inside the InSb layer at the three considered values of $V_D$. Clearly, among the three cases, an overall strongest field strength is found inside the InSb layer for $V_D = -2$ V, which should give a strong SOI. On the contrast, at $V_D = 11$ V, the field strength inside the InSb layer is overall small, which should produce a relatively weak SOI. Figure 3(c) displays the calculated carrier density distributions in the InSb layer at the three values of $V_D$. As shown in Fig. 3(c), at $V_D = -2$ and 0 V, carriers are non-uniformly distributed and they mainly concentrated to the bottom part of the InSb layer, close to the $SiO_2$ dielectric, where strong electric fields are presented and carriers will experience a strong Rashba SOI when they move along the layer. At $V_D = 11$ V, the carrier distribution becomes less non-uniform in the InSb layer with a significant amount appearing in the middle region of the layer, where the electric field is comparably weak and the carriers would experience a weak SOI in the InSb nanosheet. All these simulated results are in a good agreement with our experimentally measured results for the SOI in the InSb nanosheet.

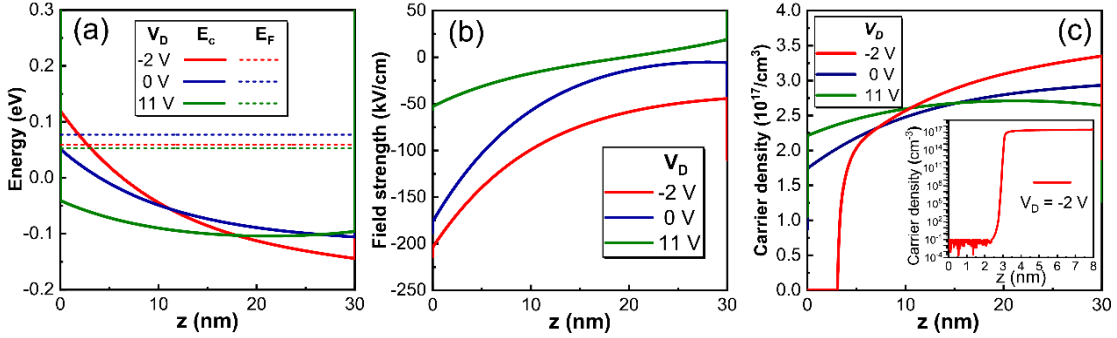

Figure 3. (a) Calculated conduction band edges (red, blue and green solid lines) and Fermi levels (red, blue and green dashed lines) inside the InSb layer with a sheet carrier density $n = 7.2 \times 10^{11}$ cm$^{-2}$ at three different values of dual-gate voltage $V_D$. (b) Calculated corresponding effective electric field strengths inside the InSb layer at the same three values of $V_D$ as in (a). (c) Calculated corresponding carrier density distributions inside the InSb layer at the same three values of $V_D$ as in (a). The inset shows the results of the calculations at $V_D = -2$ V with the carrier density plotted in logarithmic scale.

**Supplementary Note IV. Analysis of the Rashba SOI in the InSb nanosheeet**

In a semiconductor quantum structure, two predominant mechanisms that give rise to spin-orbit coupling and thus lift the spin degeneracy even in the absence of a magnetic field are the Dresselhaus and Rashba[2] SOI. The first one arises from an intrinsic bulk inversion asymmetry (BIA) of the underlying crystal structure, as described by Dresselhaus[3], while the second one arises from a structural inversion asymmetry (SIA) induced by an electrical field $\mathbf{E} = -\nabla V(\mathbf{r})$ in the crystal, where $V(\mathbf{r})$ is the electric potential, as described by Bychkov and Rashba[4]. The electric field could include both a built-in part in the structure and a tunable part created by, e.g., applying a gate voltage.

In the lowest-order approximation, the Rashba SOI Hamiltonian can be written as[5]

$$H_R = r_R \boldsymbol{\sigma} \cdot \mathbf{k} \times \mathbf{E}, \tag{3}$$

where $r_R$ is a material-specific, Fermi level dependent prefactor[6,7] and $\mathbf{k}$ is the wave vector. To estimate the effect of the Rashba SOI in our dual-gate device structure, we assume that all conduction carriers experience a same electric field in the InSb nanolayer. We approximate this electric field by the mean electric field $\mathbf{E} = (0, 0, E)$, with $E$ being the perpendicular component of the electrical field obtained by averaging through the InSb nanolayer along the perpendicular direction. The wave vector only has in-plane components and can be written as $\mathbf{k} = (k_x, k_y, 0)$. Rashba Hamiltonian then becomes $H_R = \alpha_R(k_y \sigma_x - k_x \sigma_y)$, where $\alpha_R = r_R E$ is known as the Rashba SOI strength parameter. Moreover, the spin-orbit precession length is given by $L_R = \frac{\hbar^2}{m^* \alpha_R}$ with $m^*$ being the elecron effective mass. Therefore $\frac{1}{L_R}$ is in proportion to $\alpha_R$ and thus in proportion to the mean electric field $E$.

The extracted spin-orbit relaxation length $L_{SO}$ from the measured magnetoconductance data in our experiments comprise contributions from comprehensive spin relaxation processes induced by all kinds of SOIs (i.e., Rashba SOI, Dresselhaus SOI and other high-order kinds). It is naturally a hypothesis that the Rashba SOI induced spin precession process is the major cause for the WAL characteristics observed in our magnetoconductance measurements. Thus, the spin procession induced relaxation time $\tau_{SO}$ caused by all SOIs could be written as $\frac{1}{\tau_{SO}} = \frac{1}{\tau_R} + \cdots$, where $\tau_R$ is the spin relaxation time cause by the Rashba SOI. As a consequence, we have $\frac{1}{L_{SO}^2} = \frac{1}{L_R^2} + \cdots$ and thus expect to see that $\frac{1}{L_{SO}^2} = \left(\frac{m^* r_R}{\hbar^2}\right)^2 E^2 + C_0$ in the InSb nanolayer in our device, where $C_0$ is a constant by assuming that the Elliot-Yafet term, Dresselhaus SOI term and all other high-order terms are electric field independent. To see this, we plot in Fig. 4, the extracted $\frac{1}{L_{SO}^2}$ as a function of $E^2$ at carrier densities of $n = 7.2 \times 10^{11}$ cm$^{-2}$ and $4.3 \times 10^{11}$ cm$^{-2}$. The red and blue dashed lines in the figure show the linear fits to the data at the two different carrier densities, namely different Fermi levels. As shown in Fig. 4, at both carrier densities, $\frac{1}{L_{SO}^2}$ displays a good linear dependence on $E^2$. The slopes $\kappa = \left(\frac{m^* r_R}{\hbar^2}\right)^2$ of the fitting lines are 0.613 V$^{-2}$ at $n = 7.2 \times 10^{11}$ cm$^{-2}$ and 0.41 V$^{-2}$ at $n = 4.3 \times 10^{11}$ cm$^{-2}$. This result supports our above assumption that the Rashba SOI induced spin precession process is the major cause for the observed gate voltage tunable WAL characteristics. The intercepts of the two fitting lines and the vertical axis are nearly the same and are very close to a value of 8 μm$^{-2}$, which gives $C_0 = 8$ μm$^{-2}$ and represents all other field-independent contributions including the one from the Dresselhaus SOI. The Fermi level dependent prefactors $r_R$ can be obtained from the extracted slopes $\kappa$ using the relation $r_R = \frac{\hbar^2}{m^*} \cdot \sqrt{\kappa}$. The results are $r_R = 4.26$ $e \cdot$nm$^2$ at $n = 7.2 \times 10^{11}$ cm$^{-2}$ and $r_R = 3.48$ $e \cdot$nm$^2$ at $n = 4.3 \times 10^{11}$ cm$^{-2}$.

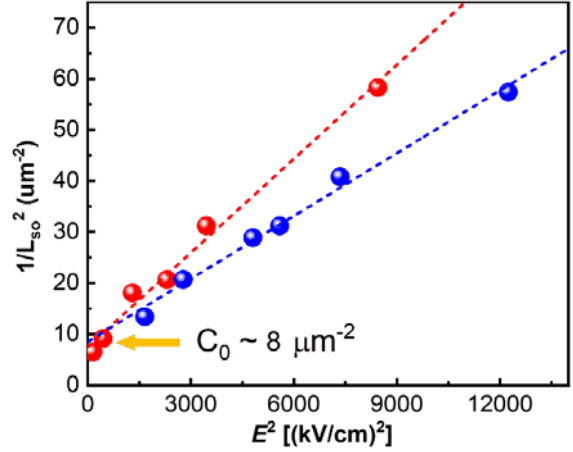

Figure 4. Extracted $\frac{1}{l_{so}^2}$ versus calculated mean field strength $E^2$ in the InSb nanolayer of the dual-gate device. Red and blue dots are the data points obtained at sheet carrier densities $n = 7.2 \times 10^{11}$ cm$^{-2}$ and $4.3 \times 10^{11}$ cm$^{-2}$, respectively. Red and blue dashed lines are the linear fits to the data. The two fitting lines intersect the vertical axis at nearly the same value of $C_0 \sim 8$ μm$^{-2}$ as marked by a yellow arrow.

**Supplementary Note V. Dual-gate voltage dependent measurements of the magnetoconductance along the constant conductance contour lines of ~1.1 and ~2.6 $e^2/h$**

The dual-gate voltage dependent measurements of the magnetoconductance and the characteristics transport lengths have also been performed for the InSb nanosheet at two lower carrier densities, i.e., along the constant conductance contour lines of ~1.1 and ~2.6 $e^2/h$. Figure 5 summarizes the measurements, where Figs. 5(a) and 5(b) show the results along the constant conductance contour line of ~2.6 $e^2/h$, while Figs. 5(c) and 5(d) show the results along the constant conductance contour line of ~1.1 $e^2/h$. It is seen that similar dual-gate voltage dependences of the transport lengths $L_\varphi$, $L_{so}$, and $L_e$ as observed in Fig. 3 of the main article are obtained. In particular, the spin-orbit length $L_{SO}$ is seen to be efficiently controlled via the dual-gate voltage $V_D$ at both constant conductance values. These data, together with those shown in Fig. 3 of the main article, demonstrate that the SOI in the InSb nanosheet in a dual-gate structure can be efficiently tuned by a voltage applied to the dual gate at largely different but fixed carrier densities of the nanosheet.

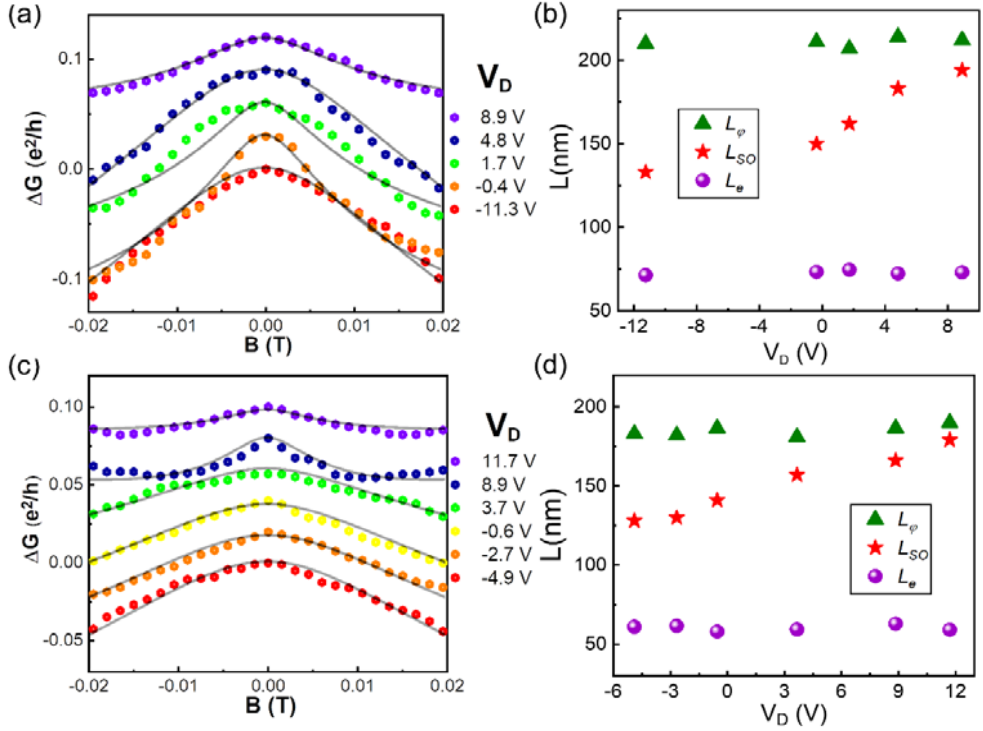

Figure 5. (a) Low-field magnetoconductance $\Delta G$ measured for the device at a constant conductance value of ~ 2.6 $e^2/h$ and temperature $T$ = 1.9 K at various values of the voltage $V_D = V_{TG} - V_{BG}$ applied over the dual gate. The bottom trace shows the magnetoconductance data measured at $V_D = -11.3$ V and all other measured magnetoconductance traces are successively vertically offset for clarity. The black solid lines are the theoretical fits of the experimental data to the HLN equation [Eq. (1) in the main article]. (b) Phase coherence length $L_\varphi$, spin-orbit length $L_{SO}$, and mean free path $L_e$ extracted from the fits in (a) as a function of $V_D$. (c) The same as (a) but measured for the device at a constant conductance value of ~1.1 $e^2/h$. Here, the bottom trace shows the magnetoconductance data measured at $V_D = -4.9$ V and all other measured magnetoconductance traces are again successively vertically offset for clarity. (d) Phase coherence length $L_\varphi$, spin-orbit length $L_{SO}$, and mean free path $L_e$ extracted from the fits in (c) as a function of $V_D$.